# Transformations in Perovskite Photovoltaics: Film Formation, Processing Conditions, and Recovery Outlook

Bidisha Nath[a], Jeykishan Kumar[a], Sushant K Behera[b], Praveen C Ramamurthy[a,b*], Debiprosad Roy Mahapatra[c], Gopalkrishna Hegde[d]

a Interdisciplinary Centre for Energy Research, Indian Institute of Science, Bengaluru-560012, Karnataka, India

b Department of Materials Engineering, Indian Institute of Science, Bengaluru-560012, Karnataka, India

c Department of Aerospace Engineering, Indian Institute of Science, Bengaluru-560012, Karnataka, India

d Centre for Biosystems Science and Engineering, Indian Institute of Science, Bengaluru-560012, Karnataka, India

## Abstract

Organometallic halide perovskites have garnered considerable attention in recent times due to their promising optoelectronic attributes, particularly within the realm of solar photovoltaics (PV). How perovskite films form is of utmost significance in shaping their structural and functional characteristics. In this context, the application of methylamine vapour during the precursor deposition and subsequent treatment during the film formation stages emerges as crucial for the development of high-quality perovskite films for solar cell applications. The utilization of methylamine vapour annealing is pivotal in improving the crystallinity, morphology, and overall integrity of perovskite films. This work investigates the characteristics of perovskite films based on methylamine lead iodide, focusing on aspects such as crystallographic structure and vibrational modes, which are directly linked to the performance of the devices. The maximum power conversion efficiencies (PCE) obtained are 19.5% and 18.6% using 1-step and 2-step processes are obtained. The effect of factors like trap states, film homogeneity, and interfaces on the device performance are explored through capacitance measurements, photoluminescence, and electroluminescence behaviour. The recombination behaviour of the perovskite films is correlated with the crystallographic properties. These findings provide valuable insights into the influence of different processing techniques, such as methylamine vapour treatment and vacuum annealing, on rejuvenating perovskite solar cells.

**Keywords**- Perovskite solar cells, film quality, fabrication condition, electroluminescence, recovery,

## Introduction

Perovskite solar cells (PSCs) are the fast-growing third-generation solar cells[1] because of the interesting optoelectronic properties of the perovskite, high charge carrier mobility, diffusion length, defect tolerance, tuneable properties, hot carrier extraction[2] and easy processing conditions, simple solution processibility of the perovskite put PSCs in the low cost-high efficiency devices. The properties of the perovskite can be altered by the change in fabrication conditions [3]. Perovskite films are typically fabricated using solution processing techniques such as spin-coating, doctor blading, or inkjet printing. These methods involve depositing a precursor solution onto a substrate and allowing it to undergo a series of chemical reactions to form the desired perovskite structure. These fabrication conditions highly affect the film quality by affecting the defect characteristics[4], trap density of states, inhomogeneity etc [5]. Factors such as processing conditions, purity and composition of materials, kinetics of nucleation and growth, and crystallisation processes critically influence the quality of films, which in turn, significantly impacts device performance.

The morphology and crystallinity of perovskite films strongly affect their charge transport, recombination rates, and light absorption properties. These properties also affect the interfacial charge transport between the active layer and the transport layer [6]. By exploring the film formation dynamics, the aim is to gain insights into the nucleation, growth, and transformation processes that govern the development of perovskite films[7]. This knowledge can guide the development of fabrication techniques to achieve controlled film morphology and enhanced crystallinity, ultimately resulting in improved device performance[8]. The thermodynamics of crystallisation dictate the transformation of the amorphous precursor into the crystalline perovskite structure. The crystallisation process is influenced by factors such as temperature, cooling rate, solvents [9] and precursor composition. The driving force for crystallisation is the reduction in free energy associated with the formation of the ordered perovskite lattice. Understanding the thermodynamics of crystallisation helps optimize the processing conditions to achieve high-quality perovskite films with desired crystal structures. To understand the role of solvent composition in the crystallisation of perovskite [10], ternary phase diagrams are also reported.

Perovskites are vulnerable to degradation caused by various factors such as light, humidity, and oxygen. To enhance the quality of perovskite films and enable the recycling of perovskite solar cells, amines like butylamine and methylamine are commonly employed [11]. Exposure to these vapours can potentially alter film quality by affecting the kinetics of film formation, a process dictated by thermodynamics. While methylamine vapour treatment has gained popularity, its extensive exploration remains limited, aiming to mitigate the need for antisolvent treatment on a large scale and rejuvenate perovskite solar cells. Despite the usefulness of amine-based vapour treatment, the mechanisms and the structure-property relationship involved in the recrystallisation of perovskite films remain unclear, necessitating further research. These film processing techniques present significant potential for improving the recuperation of perovskite-based solar devices that have reached their end of life, primarily due to the degradation of the perovskite layer.

This work delves into the examination of perovskite films and devices based on $MAPbI_3$ following exposure to distinct film processing conditions. These film processing conditions, including vacuum annealing and methylammonium (MA) vapour exposure, exert considerable influence on the crystallographic attributes and optoelectronic properties of the materials. Importantly, these processes hold potential for application in large-scale manufacturing settings, with the capability to enhance the overall quality of the produced films. The impact of these film processing conditions on the films' characteristics resonates deeply with the defect properties, charge transport dynamics, carrier accumulation, and recombination phenomena within the n-i-p structured devices. Here a comprehension of how alterations in film attributes reverberate across the device architecture. Additionally, the study encompasses an assessment of device quality vis-à-vis different processes, elucidated through photoluminescence (PL) and electroluminescence (EL) measurements. Furthermore, the study discusses the rejuvenation of moisture-degraded perovskite films using MA vapour.

**Experimental section**- Comprehensive characterizations, information about data processing software, and details of fabrication procedures are provided in the supplementary information.

## Results and discussions
**Film formation, crystallographic and morphological analysis**

Here the perovskite films are solution-processed (Fig. 1a), followed by various film processing conditions like vacuum annealing, vapour treatment etc. Nucleation, growth, dissolution, and recrystallisation constitute critical stages in the formation of perovskite films, as the morphology of the active layer significantly governs the electronic performance of the device [12]. Some abbreviations are used in the manuscript indicating the fabrication conditions of the films- MA- Methylamine vapour treated, VA- Vacuum annealed, MAVA- Methylamine vapour treatment followed by vacuum annealing, 1 and 2 indicating one-step and two-step fabrication conditions.

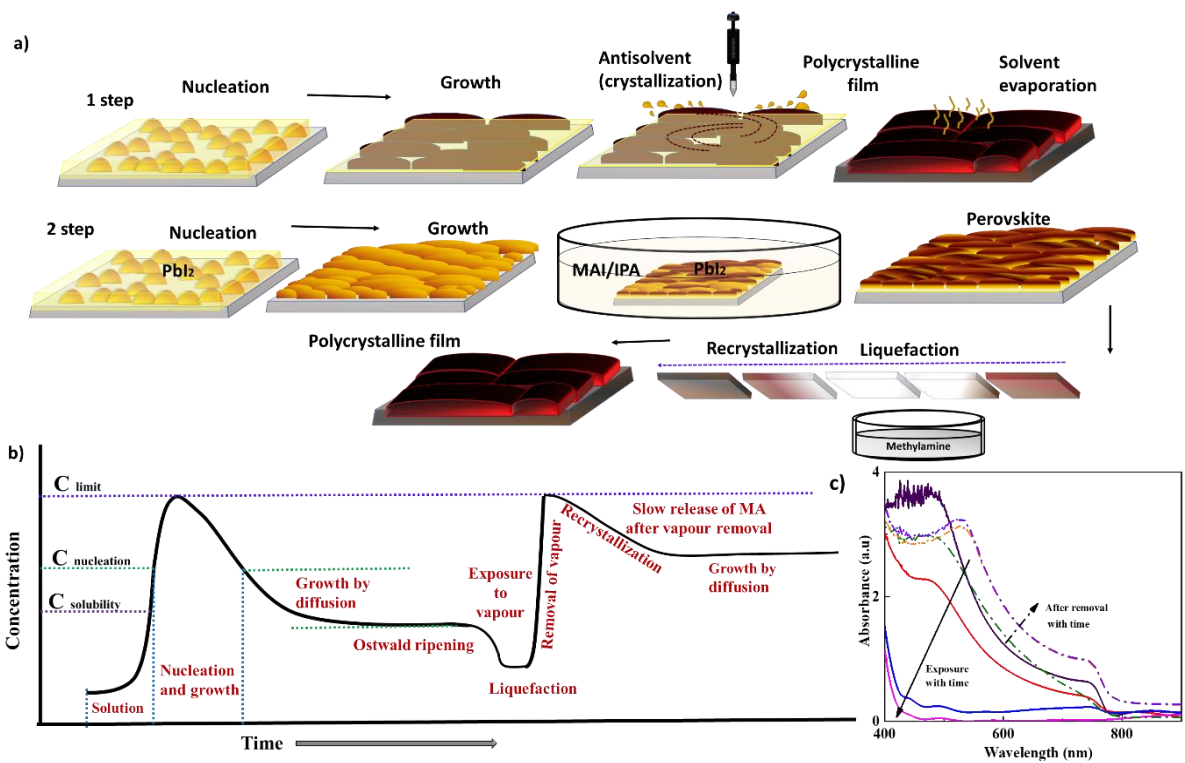

Fig.1 a) Fabrication process, b) Proposed curve for perovskite film growth and recrystallisation after MA vapour exposure, and c) In-situ absorption spectra with the MA vapour exposure and after its removal

Nucleation is influenced by factors such as solution concentration, temperature, and the presence of nucleation agents. Once nucleation occurs, the crystals grow through the diffusion and deposition of additional precursor molecules onto the existing crystal surfaces. The growth process can be affected by parameters such as solution composition, temperature, and annealing conditions. The Lamer's curve for film formation [13] and MA recrystallisation for perovskite is proposed in Fig.1b. (Lamer's curve for spin-coated perovskite in Fig.S1)

Based on the ΔH (enthalpy) values for film formation [14], a comparative table is generated to assess the thermodynamic stability (Table S1). While annealing, the wet film gets enough activation energy to break the precursor-solvent complexes. With increasing temperature, $C_p$ increases and thereby H increases. The enthalpy of formation ($\Delta H_{form}$) for perovskite film is higher than that of the solution. As the temperature increases, the $T\Delta S$ part becomes more negative (T= temperature, $\Delta S$= entropy), lowering the Gibbs free energy ($\Delta G$), resulting in the formation of perovskite film. When the $MAPbI_3$ perovskite solution is deposited and forms a solid film, the solute molecules come into closer proximity and start to interact more strongly with each other. This can lead to the formation of a well-ordered crystalline structure within the film. As a result, the free energy of the $MAPbI_3$ perovskite film is typically lower than that of the dissolved solution. In the antisolvent treatment process, solvent removal occurs rapidly and is optimized during spinning, leading to a quicker attainment of the critical Gibbs free energy ($\Delta G_c$), thereby reaching the critical radius of nucleation ($r_c$) faster compared to the perovskite crystallized without involving antisolvent [15].

Within the framework of methylammonium (MA) vapour liquefaction and recrystallisation, the formation enthalpy ($\Delta H_{form}$) of the $MAPbI_3 \cdot MA$ complex is observed to be higher than that of solid $MAPbI_3$, closely mirroring the properties of the $MAPbI_3$ solution (Table S1). This indicates that the perovskite film experiences liquefaction as a result of MA vapour absorption, without the need for external heating. However, this transparent complex exhibit instability; upon the removal of the vapour, the film reverts to its original dark brown colour. The adsorption energy for MA vapour can vary from ~ 101.3 kJ/mol to ~ 67.54 kJ/mol [16]. The driving forces for the MAI adsorption in the perovskite film inhomogeneity i.e. Pb rich or MA rich conditions, surface defects, dangling bonds etc. $\Delta S$ in perovskite consists of two parts- $\Delta S_{vibrational}$ and $\Delta S_{configurational}$. The vibrational entropy can be correlated with the phonon modes of the perovskites After the adsorption of MA vapour, the $\Delta S$ increases, without temperature change. In this case, the configurational entropy is expected to dominate the vibrational entropy as there is no annealing or rise in temperature. The vibrational entropy only comes from the room-temperature phonons. The introduction of methylamine gas can disrupt the equilibrium within the $MAPbI_3$ crystal lattice, leading to a redistribution of MA cations and a rearrangement of the crystal structure. This rearrangement may increase the number of possible configurations or orientations of the MA cations, thereby increasing the configurational entropy. Annealing is done after the film turns brown to achieve a higher grain size and compact films. Upon removal of the methylamine gas, the perovskite lattice tends to return to its original state, with the MA cations assuming their preferred positions. This relaxation process reduces the overall energy of the system, resulting in lower free energy.

When untreated perovskite films were exposed to methylamine vapour, a noticeable change in their absorption behaviour was observed, as shown in Fig. 1c (Absorbance behaviour for all different films in Fig. S2a). Upon exposure to the vapour, the films became transparent, leading to a decrease in the absorption spectra. Once the vapour was removed, the perovskite films gradually reverted to their original absorption behaviour. The reduction in the absorption spectra can be attributed to the liquefaction of the perovskite film and the intercalation of MA. Notably, the absorption spectrum in the presence of MA vapour differs significantly from the absorption behaviour of $PbI_2$. Fig. S2b and S2c display the absorption behaviour of $PbI_2$ films, and the solvents used, including perovskite solutions and perovskite in methylamine, highlighting that the kinetics of perovskite in DMF-DMSO solution and in methylamine diverge from those observed under solid film conditions. Huang et al. reported similar absorbance properties of this intermediate phase, i.e. $CH_3NH_2\cdots CH_3NH_3^+$ dimers along with $PbI_3^-$ dimers [17]. In the one-step $MAPbI_3$ coating process, film formation follows the classic Lamer curve with three stages: solution, nucleation and growth, and growth. Nucleation during the formation of perovskite films tends to be relatively slow, which typically leads to the formation of larger grains, reduced surface coverage, and an increase in the number of grain boundaries. This characteristic can impact the film's uniformity and electronic properties. To remove any residual solvent, annealing is performed after the spin coating process. Films produced via a one-step spin-coating technique often lack texturing and might contain residual $PbI_2$, which can adversely affect the device's performance. In contrast, the two-step perovskite coating process begins with the deposition of a $PbI_2$ layer via spin coating. The solvent DMF is used to dissolve $PbI_2$, with DMSO added as an additive to achieve the desired morphology crucial for high-performance devices. This $PbI_2$ film is then converted to $MAPbI_3$ through a reaction with MAI, where MAI intercalates within the $PbI_2$ lattice. This intercalation process results in an overall volume expansion of the material. However, during this intercalation process, the lack of control can lead to numerous grains protruding from the surface, which contributes to the roughness of the resulting film (Fig. S3). This roughness will influence the film's optoelectronic properties and, by extension, the performance of the perovskite-based device.

Perovskite films fabricated using two-step process, have shown texturing in the plane like $PbI_2$ [18]. Texturing is observed in two-step films, along with diffraction rings [19,20] related to residual $PbI_2$. Incomplete conversion, volume expansion, and lattice strain hinder high-quality film growth. $PbI_2$ films formed solely with DMF undergo dissolution-recrystallisation and volume shrinkage after DMF removal through annealing. DMSO added to the solution induces solvent annealing, and coarsening $PbI_2$ grains. DMSO forms a stronger complex with $PbI_2$, pre-expanding the lattice before MAI intercalation, reducing strain and micro-cracks in the perovskite film. Compared to one-step films, two-step process grains are smaller and energetically less stable due to their higher chemical potential values [21]. The solvent in the perovskite film evaporates rapidly due to the swift solvent removal, causing the wet film on the substrate to reach supersaturation almost immediately. This condition leads to quick nucleation, which is anticipated to produce compact perovskite films with reduced voids and improved packing density. Despite this, no significant crystallographic differences are observed between one-step and two-step vacuum annealed films, as evidenced by grazing incidence wide-angle X-ray scattering (GIWAXS) data presented in Fig. 2. This suggests that merely vacuum annealing the perovskite films, without any specific solvent or vapour treatment, does not substantially enhance film quality; in fact, it tends to result in coarser perovskite grains. Interestingly, vacuum annealing post-methylamine treatment appears to improve the orderliness within the films for both one-step and two-step processed films, indicating that this combined treatment can positively impact the film structure. The perovskite solution preparation involves the use of DMF and DMSO as solvents. Antisolvent treatment is specifically employed to rapidly remove DMF from the perovskite film, while residual DMSO is eliminated through thermal annealing. The rate at which these solvents are removed critically influences the morphology of the perovskite grains, underscoring the importance of solvent management in achieving optimal film characteristics for perovskite-based devices. The concentration of the film remains above the critical concentration threshold, prompting the formation of new nuclei. The initial transparency of the film can be attributed to the presence of DMSO in the solution, which delays the reaction between MAI and lead iodide $PbI_2$. This delay is due to the formation of a strong complex between $PbI_2$ and DMSO. During the process, DMSO is gradually removed from the $MAI:PbI_2:DMSO$ intermediate through annealing, allowing the reaction to proceed and the perovskite structure to form more fully. This mechanism highlights the critical role of DMSO not only in the initial stages of film formation but also in the subsequent thermal treatment, which is essential for achieving the desired perovskite film quality and morphology [22]. The $MAPbI_3$ forming on $PbI_2$ in 2-step processes are showing face-up and corner-up GIWAXS-based features. The face-up feature gets more prominent with the increase in grazing angle (Supporting information Fig. S6).

The crystallographic texturing effect is observed after the exposure of methylamine vapour (Fig. 2). Crystallographic texturing can be the cause of substrate properties, strain, minimization of surface energy, additives and templating agents, coating environment, heterogeneity in the film etc. Exposure to methylamine vapour causes the formation of a liquid-state intermediate, and then recrystallisation and thus resulting random orientation of the grains [23] (Table S2). This mechanism is studied by Zhou et al. [16]. Considering LaMer's curve for film formation, the perovskite films liquify in the presence of MA vapour i.e. the solution condition. Upon removal of the vapour, supersaturation is obtained. Nucleation and growth start. Slow release of excess $MA^+$ from the lattice leads to the formation of large grains. The established phenomenon of methylamine (MA) molecule interaction with the inorganic $[PbI_6]^{4-}$ framework is rooted in the coordination between the unshared electron pairs present on the nitrogen atoms within methylamine and the vacant orbitals associated with $Pb^{2+}$ cations. In accordance with documented research outcomes, the transition from the three-dimensional (3D) α phase to the one-dimensional (1D) δ phase, a process integral to liquefaction, is instigated by the formation of $MA^-MA^+$ dimers [24]. The methylamine molecule establishes a robust covalent linkage with the Pb atom, resulting in a binding energy of 1.1 eV. This robust linkage effectively hinders the migration of methylamine gas across the perovskite surface and its dissolution on the $PbI_2$-terminated surface when exposed to methylamine gas. On the MAI-terminated surface, the energy required for the formation of a methylamine vacancy is notably low, measuring at 0.5 eV, which implies a probable presence of such vacancies within perovskite structures. Adsorbed methylamines exhibit a discernible preference for infiltrating the defective MAI-terminated surface over the pristine surface, suggesting that the dissolution of perovskite thin films in methylamine vapour may be facilitated by surface processes assisted by defects [16].

For 2-step MA films highly oriented perovskite grains i.e. grains with a preferable orientation at the surface lead to the distinctive diffraction dots in the GIWAXS pattern. As the penetration angle increases, dots get connected to give a fairly continuous ring with non-uniform intensity which can be indicative towards the increase in polycrystalline nature and slight orientation on the preferred planes while going away from the surface (A detailed chart of angles related to grain orientation is provided in the supporting information Table S2). For the 1-step MAVA and 2-step MAVA films, surface texturing is observed from the GIWAXS rings. As the depth increases from the surface this texturing effect reduces, giving rise to continuous rings with features which are commonly observed in case of the presence of a phase mixture. A combination of 2D and 3D phases have been reported to show such Debye–Scherrer rings [25]. Exposure to methylamine vapour may give rise to MA excess condition in the films. In the presence of a high concentration of excess MAI, the basic $(PbI_6)^{4-}$ units are found as individual zero-dimensional (0D) quantum dots within an MAI matrix. When the quantity of MAI is reduced, these 0D units form connections and gradually construct one-dimensional (1D) perovskite chains and two-dimensional (2D) perovskite sheets through self-assembly. This process ultimately results in the formation of three-dimensional (3D) perovskite structures [26]. Under such conditions, a combination of 3D and lower-dimensional perovskites may arise [27]. This phenomenon may be attributed to the methylammonium-methylamine dimer, which constitutes a relatively large cation, expected to facilitate the formation of low-dimensional perovskite structures [28]. The perovskite films, without any optimized antisolvent treatment, demonstrate improved quality following exposure to methylamine vapour. Perovskite grains exhibiting a compact morphology are observed, as confirmed by morphological analysis. Films initially exhibiting superior quality deteriorated after exposure to MA vapour (Supporting Information Fig. S4). Specifically, when the antisolvent-treated one-step glossy dark perovskite film was subjected to MA vapour, a decrease in grain size occurred, leading to a non-uniform and rough surface characterized by protruding grains. To facilitate a comparison of the crystallographic data for all the films, the results are summarized in Table 1.

GIWAXS [29,30] gives information about the crystallographic planes with different orientations along with parallel to the substrate ($q_z$) and planes perpendicular to the substrate ($q_{xy}$) i.e. out of plane and in-plane scattering. All the perovskite films obtained from different processes show the highest ordering considering the 110 plane at q=~1Å$^{-1}$ [31]. The Debye–Scherrer rings is most prominent for 110 planes. At GI 0.1, the nature of the peak, as well as diffused rings from GIWAXS patterns, can be attributed to surface defects and non-uniformity (Supporting information Fig. S5, S6, S7). At a higher grazing angle, the rings and the peaks appears to become sharper. Diffraction patterns obtained from the in-plane and out-off-plane line scan patterns are plotted in Fig. 3. For the 1-step coated films (1 step, 1 step-VA and 1 step AS), the orientation of planes is very similar for both parallel and perpendicular conditions. For 2 step-coated films (2-step, 2-step-VA and 2-step AS), the orientation (110) is prominent considering out of plane scattering ($q_z$) condition. Considering $q_{xy}$ the lead iodide peaks ((001) and (101)) are more prominent [32]. The peak for 110 and 220 planes are almost overlapping for in-plane and out-of-plane conditions for 1-step, 1-step-VA and 1-step AS films. Residual $PbI_2$ peaks of 001 and 101 are observed in two-step films except 2-step MAVA. The $q_z$ and $q_{xy}$ patterns are also matching for 1-step MAVA and 2-step MAVA. The vacuum annealing improves the film quality by reducing the lattice strain. After the MA vapour treatment, the peak at 110, and 220 become prominent [17]. This might be the effect of atomic arrangement in the unit cell [20].

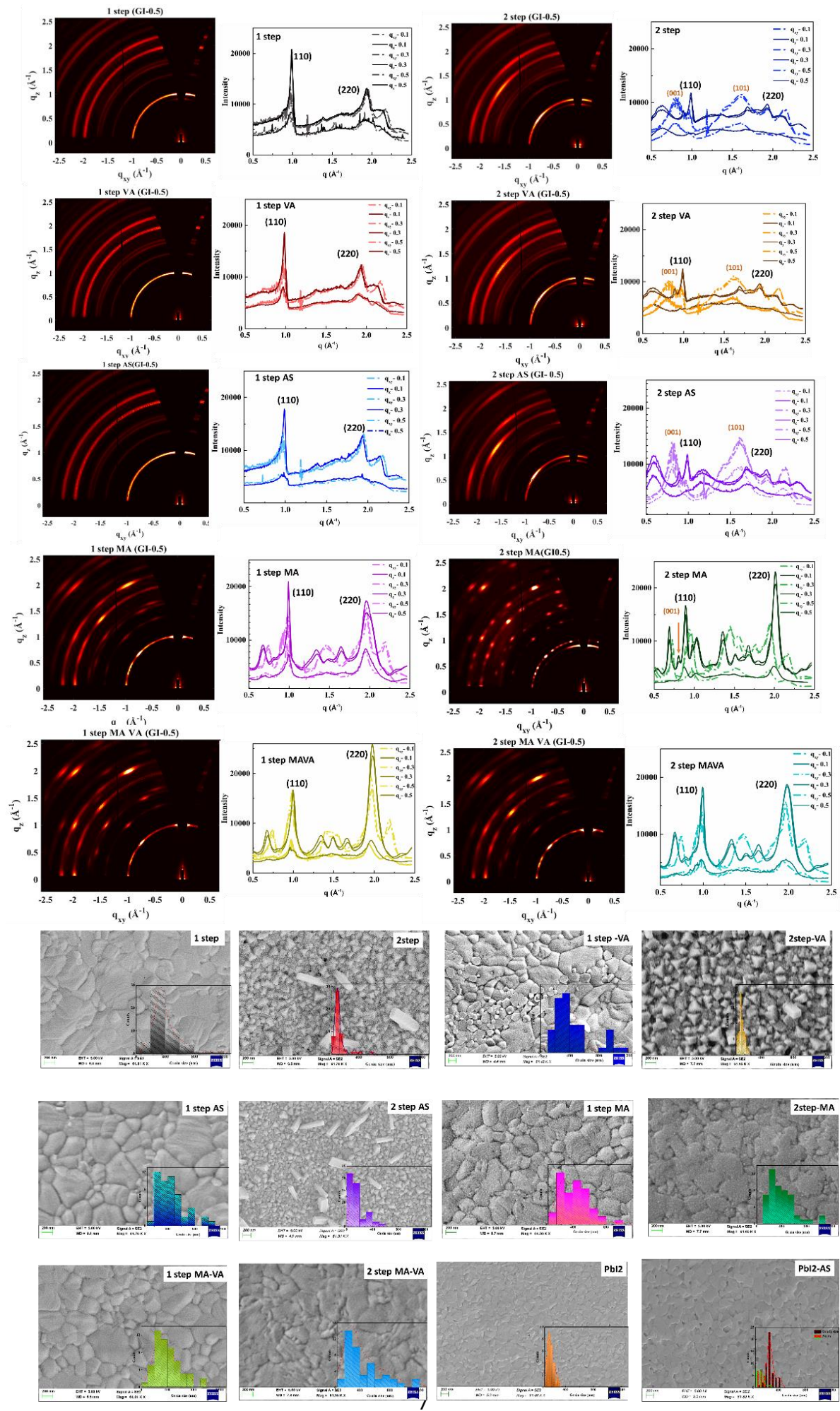

Fig. 2 GIWAXS and corresponding line scan (in-plane, out-off-plane pattern for perovskite films with different fabrication processes with and without antisolvent treatments (AS) and post-fabrication treatments like vacuum annealing (VA), Methylamine (MA) vapour exposure and Methylamine- vacuum annealing (MAVA); Morphological information or the perovskite films

Table 1 Inference of the crystallographic measurements-table

| Films | Debye–Scherrer rings | Preferred Orientation Planes | In-plane and out off plane | Inference |
|---|---|---|---|---|
| **1 step** | Continuous | 110, 220 | Matching | Polycrystalline, fine grains |
| **2 step** | Continuous | 110, 220, 001 ($PbI_2$), 101 ($PbI_2$) | In-plane- $PbI_2$ prominent, out off plane- perovskite peak prominent | Polycrystalline, fine grains |
| **1 step VA** | Continuous | 110, 220 | Matching | Polycrystalline, fine grains |
| **2 step VA** | Continuous | 110, 220, 001 ($PbI_2$), 101 ($PbI_2$) | In-plane- $PbI_2$ prominent, out off plane- perovskite peak prominent | Polycrystalline, fine grains |
| **1 step AS (ref)** | Continuous | 110, 220 | Matching | Polycrystalline, fine grains |
| **2 step AS** | Continuous | 110, 220 | In-plane- $PbI_2$ prominent, out off plane- perovskite peak prominent | Polycrystalline, fine grains |
| **1 step MA** | Continuous ring with non-uniform intensity | 110, 220 | Matching | Slight preference for orientation |
| **2 step MA** | Spots- coarse grains | 110, 220, 001 ($PbI_2$) | Micro-strain | High preference of orientation, highly intense 220- probably because of the different atomic arrangement i.e. presence of heavy element |
| **1 step MAVA** | Continuous ring with non-uniform intensity | 110, 220 | Matching | Slight preference for orientation, Highly intense 220- probably because of the different atomic arrangement i.e. presence of heavy element |
| **2 step MAVA** | Continuous ring with non-uniform intensity | 110, 220 | Matching | Slight preference for orientation |

**Analysis of the phonon modes**

An examination of phonon modes is conducted to elucidate the level of molecular disorder present in the films. In Fig. S8, the Raman spectra of the film with peaks fitted using Voigt fit and analysed. The Pb-I cage represents the collective movements of the lead (Pb) and iodine (I) atoms, whereas the librational modes of the MA cation pertain to the oscillatory motions of the entire MA molecule within the crystal structure. Anharmonicity and Vander Waals force effects are reported to be less significant in the Raman characteristics [33]. The discernible disparities observed within these two wavenumber regions can be ascribed to the divergence in the fundamental physical mechanisms and energy landscape inherent to each group of modes. This disparity is principally driven by the formation of substantial hydrogen bonds between the hydrogen atoms in methylamine (MA) and the iodine atoms present in the $PbI_6$ perovskite octahedra. Consequently, the Raman activity of MA demonstrates heightened responsiveness to structural perturbations occurring within the inorganic framework. The MA torsional modes in the higher wavenumber range are more sensitive to the local environment surrounding the MA cation and the interactions between neighbouring molecules [34]. Besides this, these corner-based MA cations show several additional degrees of rotational and translational freedom. This may give rise to the disorder in the system. The torsional mode is reported to have a direct connection with the interaction between the motion of cations and the Pb-I octahedra. The blue shift of the peak at ~215

cm$^{-1}$ can be related to the distortion of the octahedra [35]. Discrepancies in phonon modes become evident when variations in fabrication processes are introduced, even though the material itself remains consistent (detailed data provided in the supporting information Table S3). Phonon modes exhibit sensitivity to vibrational entropy, a parameter typically influenced by factors such as atomic positions, bond strength, atomic mass, and electronegativity, among others. This, in turn, indirectly influences configurational entropy. As indicated in the table (Table S1), it is noteworthy that the entropy of the MAPbI$_3$-MA complex surpasses that of MAPbI$_3$. Consequently, one can anticipate that even subsequent to the removal of MA vapour, the films will retain higher entropy in comparison to the reference material. The peak splitting serves as an indicative marker of lattice strain. Specifically, the two-step film exhibits peak splitting in the approximate range of 143-135 cm$^{-1}$, corresponding to the presence of MA$^+$ ions. Furthermore, substantial distinctions are also discernible in the q-range graphs depicting in-phase and out-of-phase scattering. Following MA vapour exposure and vacuum annealing, a pronounced increase in the contributions of MA cations is observed.

**Morphological analysis**

Film processing plays a pivotal role in modulating the morphology of perovskite films, a critical determinant of the efficiency of PSCs as shown in Fig. 4. The resulting morphology exerts control over key characteristics such as crystalline structure, grain dimensions, defect concentration, and overall film quality. A technique known as antisolvent treatment involves the deposition of the precursor solution, followed by rapid immersion in a non-solvent medium. This method yields highly defined and uniform perovskite films. In contrast, untreated films produced through a simplified single-step process exhibit non-uniformity, incomplete precursor conversion, and impaired charge mobility, all of which exert adverse effects on device performance. Vacuum annealing, a widely employed process, serves to enhance the morphology and crystalline properties of thin films, including those of perovskite. This procedure enables controlled crystallisation of the perovskite film, leading to improved packing of its constituent grains. Consequently, the resulting film structure becomes more compact and denser. In the context of two-step processes, the morphology of the perovskite film is markedly influenced by the underlying PbI$_2$ layer. Following antisolvent treatment, the PbI$_2$ layer demonstrates a porous nature, which facilitates interdiffusion of the precursor materials and enhances their conversion into perovskite. However, this yields a grain size distribution skewed towards smaller grains. While smaller grains are known to exhibit lower internal strain than larger grains, they are also associated with a higher density of grain boundaries in the film, leading to increased losses of charge carriers. Exposure to methylamine vapour induces recrystallisation in the perovskite film. Both one-step and two-step fabrication processes exhibit improved film morphology following such treatment. Furthermore, incorporating vacuum annealing subsequent to MA vapour exposure promotes the growth of densely packed perovskite films. This sequential enhancement of film characteristics contributes to the augmented power conversion efficiency (PCE) of the resulting solar devices. The surface texture of the absorber layer holds a significant influence on the operational efficiency of the device. An analysis of root mean square (RMS) roughness was performed on the films using atomic force microscopy (AFM) images, as detailed in the supplementary information. Results indicate that films subjected to MA vapour treatment exhibit a surface roughness of less than 10 nm. In contrast, the single-step AS film displays a roughness of approximately 11 nm. Conversely, the 1-step, 2-step, 1-step VA, 2-step VA, and 2-step AS films all exhibit markedly higher levels of roughness (Fig. S9). This heightened roughness significantly obstructs the effective interfacial charge transport within the devices.

**Device performance correlation with film characteristics**

In the next section, the device performance is analysed considering the current-voltage and capacitive behaviour. 2 step MA-VA-based device has shown PCE almost in the similar range to the device fabricated using conventional 1-step involving antisolvent wash (Table 2). Exposure to MA has caused the recrystallisation resolving the issue of incomplete conversion of the PbI$_2$. Vacuum annealing treatment to these films is expected to help in improving the compactness of the perovskite films. Though the 110 and 220 planes belong to the same family of crystallographic planes, the higher ordering at 110 and 220 planes has been shown by the 2-step MAVA films. 1-step, 2-step, 1-step VA, 2-step VA and 2-step AS has shown poor performance which may be attributed to the poor film quality and incomplete conversion. This is because the porous nature of $PbI_2$ helps the MAI to diffuse and form perovskite. Besides this high film roughness has caused interfacial mismatch thus resulting in charge accumulation. It is already reported that the films with a preferential orientation of the long axis of the tetragonal unit cell parallel to the substrate achieve the highest short circuit currents and correspondingly the highest photovoltaic performance [36]. Though most of the films have shown the highest orientation to plane 110, the highest current is obtained for 1-step AS film followed by 2-step MAVA. One of the major causes of the lower $J_{sc}$ values for other films is a recombination of the charge carriers at the defects because of the poor film quality (Fig. 3a).

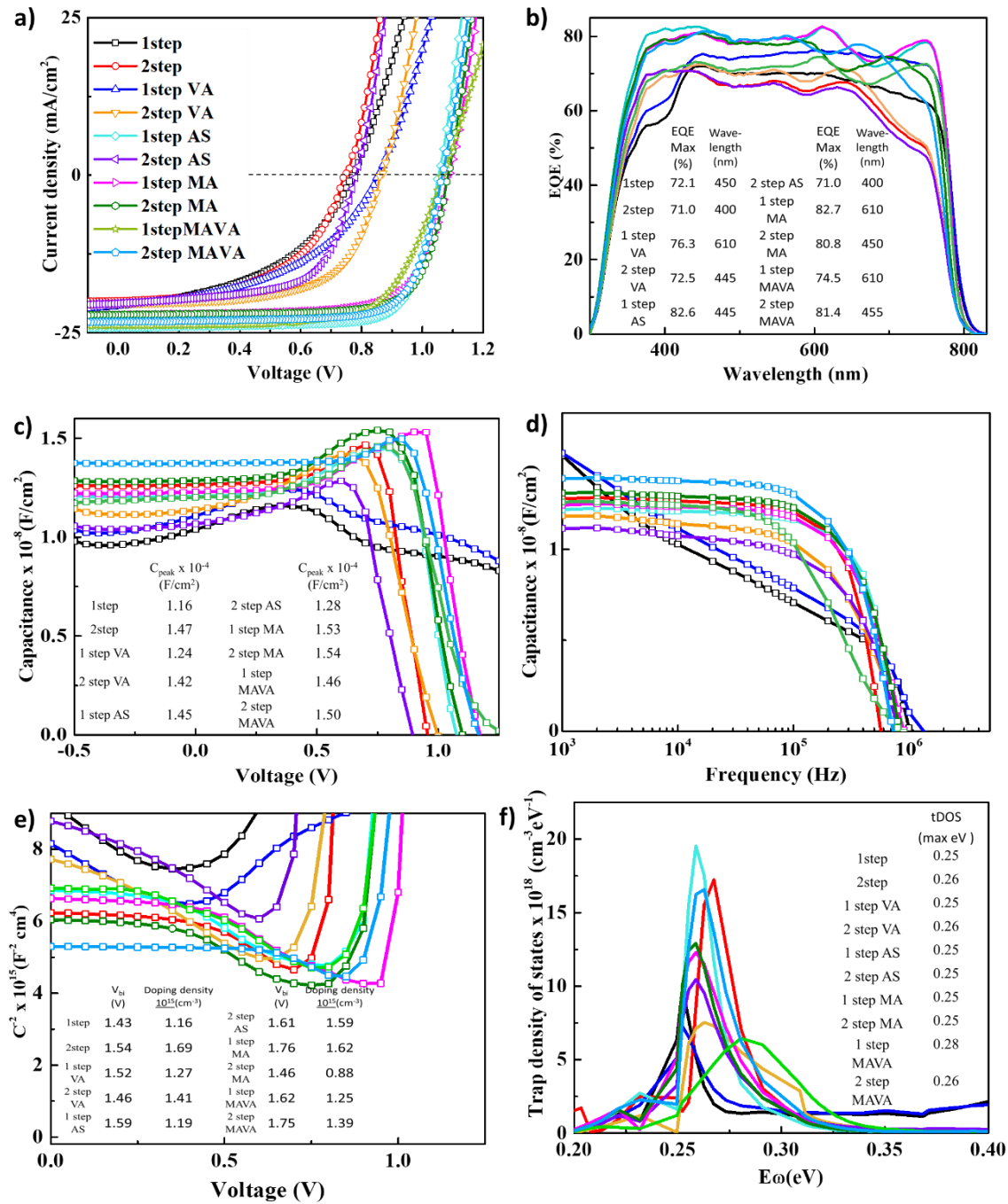

Fig. 3 Device characterization involving perovskite films with various fabrication conditions (a)Current density-voltage b)EQE and c)capacitance-voltage d) Capacitance frequency e) Mott-Schottky and f) trap density of states calculations)

Fig. 3b illustrates the external quantum efficiency (EQE) spectra of the PSCs. The formation process of the perovskite film within a photovoltaic device exerts a substantial influence on the device's EQE. To achieve elevated EQE in perovskite-based devices, the creation of a high-quality perovskite film characterized by optimal crystallinity, thickness, purity, and appropriate interfaces[37] stands as a requisite. Consistency in device performance hinges upon the uniform deposition of the perovskite film across the device area. Departures from uniformity can lead to spatial fluctuations in EQE, culminating in an overarching efficiency reduction, as evidenced in the cases of the 2-step, 2-step VA, and 2-step AS processes. This phenomenon is believed to be attributable to the presence of residual $PbI_2$, as discerned from the $q_{xy}$ and $q_z$ patterns in Fig. 2. The EQE decrement is notably pronounced on the higher wavelength side, stemming from interfacial incongruities stemming from the heightened roughness of the absorber layer within the 1-step, 2-step, 1-step VA, 2-step VA, and 2-step AS films. The film processing conditions can serve to passivate defects within the perovskite film, consequently enhancing charge carrier lifetimes and mitigating non-radiative recombination.

Table 2 Device performance

| Device | Voc (V) | Jsc (mA/cm$^2$) | FF (%) | Efficiency (%) | Rs (Ω cm$^2$) | Rsh x10$^3$ (Ω cm$^2$) |
|---|---|---|---|---|---|---|
| 1step | 0.77 | 20.65 | 46.9 | 7.4 | 7.4 | 1.2 |
| 2step | 0.75 | 19.88 | 52.8 | 7.8 | 5.5 | 2.1 |
| 1 step VA | 0.85 | 20.72 | 47.3 | 8.3 | 7.4 | 1.5 |
| 2 step VA | 0.84 | 21.25 | 59.6 | 10.6 | 4.1 | 5.9 |
| 1 step AS (ref) | 1.05 | 24.05 | 77.0 | 19.5 | 2.8 | 15.1 |
| 2 step AS | 0.81 | 20.16 | 59.3 | 9.7 | 3.7 | 2.6 |
| 1 step MA | 1.09 | 22.19 | 71.5 | 17.3 | 3.6 | 12.4 |
| 2 step MA | 1.08 | 22.13 | 75.4 | 18.0 | 3.1 | 17.5 |
| 1 step MAVA | 1.06 | 23.88 | 69.1 | 17.5 | 6.7 | 13.3 |
| 2 step MAVA | 1.07 | 23.19 | 76.1 | 18.6 | 3.2 | 16.8 |

These effects collectively contribute to the augmentation of the device's EQE. Elevated EQE values are achieved through film treatment involving either antisolvent or methylamine vapour. Further EQE losses manifest predominantly on the higher-energy flank in the context of the 1-step and 1-step VA devices. Contrastingly, for the 2-step and 2-step VA configurations, EQE reduction is prominent on the lower-energy end. This behaviour suggests suboptimal interface quality between $SnO_2$ and the perovskite in the 1-step and 1-step VA devices, while a subpar perovskite-Spiro-OMe TAD interface characterizes the 2-step and 2-step VA devices. EQE spectra exhibit similar ranges for the 1-step AS, 1,2-MA, 1-MA, and 2-MA-VA processes. The formation of the perovskite film plays a crucial role in determining the capacitive responses (Fig. 3c and 3d and Fig. S10a and S10b). The formation of a high-quality and uniform perovskite film is essential for efficient charge transport in the solar cell. A well-formed film with fewer defects and grain boundaries allows for better charge carrier mobility and reduces recombination, leading to higher capacitance. During the perovskite film formation process, some trap states may emerge within the film. These trap states can capture and hold charge carriers, leading to increased recombination rates and reduced capacitive responses. To mitigate this, the fabrication process should be optimized to minimize trap state formation. The size and orientation of perovskite grains within the film can affect the capacitive responses. Larger grains with favourable crystallographic orientation facilitate charge transport and reduce recombination, thereby improving capacitance. The capacitance-voltage measurements involve a DC bias which continuously drifts the mobile ions. The CV peak shift is mostly caused by ion migration and accumulation. The nature of the CV curve is affected by the film quality. A C-V curve for PSCs with an increase in the applied voltage shows a flat region corresponding to the depletion region, followed by an accumulation peak and finally, the junction capacitance diminishes causing carrier recombination. Usually in the PSCs, after the $C_{peak}$ shows a steep fall in capacitance. 1-step and 1-step VA-based devices have a gradual decrease in the slope while recombination of the charge carriers which is an indication of poor charge transport. The doping densities measured for all the films fall within a similar range of approximately $10^{15}$ [38]. The behaviour of capacitance is significantly influenced by both the depletion width and the barrier potential. Fig.3e shows the Mott-Schottky analysis to obtain the built-in voltage as well as the doping density. The built-in voltage shows a variation between 1.43 to 1.75 V (Fig. 3e). $V_{bi}$ is a crucial factor that influences charge collection loss. When optimized charge transport layers are present, a lower built-in potential may lead to a reversed electric field within the perovskite active layer during operation. This could result in increased recombination of the photogenerated charge carriers within the said layer [39]. The observed difference between $V_{bi}$ and $V_{oc}$ affects for efficient charge transport. A standard C-V curve comprises three regions: depletion, accumulation, and recombination. The capacitance of the depletion region is associated with the fully depleted active layer. This behaviour remains consistent until a certain voltage, after which the accumulation regime begins, leading to an increase in capacitance. This increase in capacitance corresponds to a reduction in the width of the depletion layer, until the device surpasses the barrier potential, resulting in charge recombination. Higher capacitance corresponds to a narrower depletion width. The calculation of the trap density of states (tDOS) was performed using the C-F values and tDOS is spread over an energy range of 0.2 to 0.35 eV (Fig.3f). The devices with perovskite films prepared without treatment or vacuum annealing has shown a wide spread of the tDOS, While the device with efficiencies 17-19% have a similar nature of the tDOS.

**Correlating luminescence measurements with film quality and device performance**

Considering the ease of measurements in practical scenarios, electroluminescence is a common technique for silicon solar cells to assess the quality of the solar cell. Fig. 4a and 4b, 4c show the PL and EL characteristics of perovskite films. As the perovskite material used in the devices are same, the bandgap observed from PL has very similar values (~1.6 eV) for all except 1 step and 1step VA. These films showed a red-shifted PL peak which is an indication of the presence of energy states in the forbidden energy gap. Considering the electroluminescence measurements, the injected current and voltage are lower for the high-efficiency devices. The correlation between crystallographic plane orientation and electroluminescence property can be complex and is highly material-dependent. It requires a deep understanding of both the crystallography and the underlying optoelectronic processes in the $MAPbI_3$. The influence of lattice strain and disorder on the band structure [40], subsequently impacting EL, is a well-recognized phenomenon. The presence of complete Debye-Scherrer rings serves as an indicative manifestation of crystallographic anisotropy. Given the disparity observed in the tetragonal crystal structure of $MAPbI_3$ across distinct directions, a corresponding anisotropy emerges within its optical and electronic properties, consequently imparting distinctive traits to semiconductor device performance. In the context of $MAPbI_3$, singularity prevails in carrier diffusion and recombination due to the crystal's inherent anisotropic nature. In $MAPbI_3$ perovskite, the arrangement of atoms along different crystallographic planes can influence the behaviour of charge carriers (electrons and holes) and their recombination dynamics. Some crystallographic planes might provide more favourable conditions for band-to-band recombination, resulting in more efficient photoluminescence and electroluminescence. A strategic modulation of the ratios between $PbI_2$ and MAI facilitates the simultaneous exposure of the (110), (112), (100), and (001) crystallographic planes within the $MAPbI_3$ crystal. A notable outcome of this manipulation is the heightened exposure of MAI along the (110) and (001) planes, precipitating a red shift in PL behaviour. Additionally, extended diffusion lengths and augmented carrier lifetimes manifest on these planes relative to the (100) and (112) counterparts [41]. Modulating the crystallographic orientation of perovskite films holds the potential to exert control over trap state distribution and grain boundary characteristics, consequently exerting an influence on the dynamics of charge carrier recombination. The utilization of EL and PL measurements is currently shedding light on the intricacies of band-to-band recombination processes, taking into account the quality of the film and its underlying crystallographic orientations. A comparative study in Table 3 shows that the perovskite films devoid of strain-induced effects, as well as films characterized by a pronounced alignment of crystalline grains, have demonstrated elevated external quantum efficiency (EQE_EL) values [42].

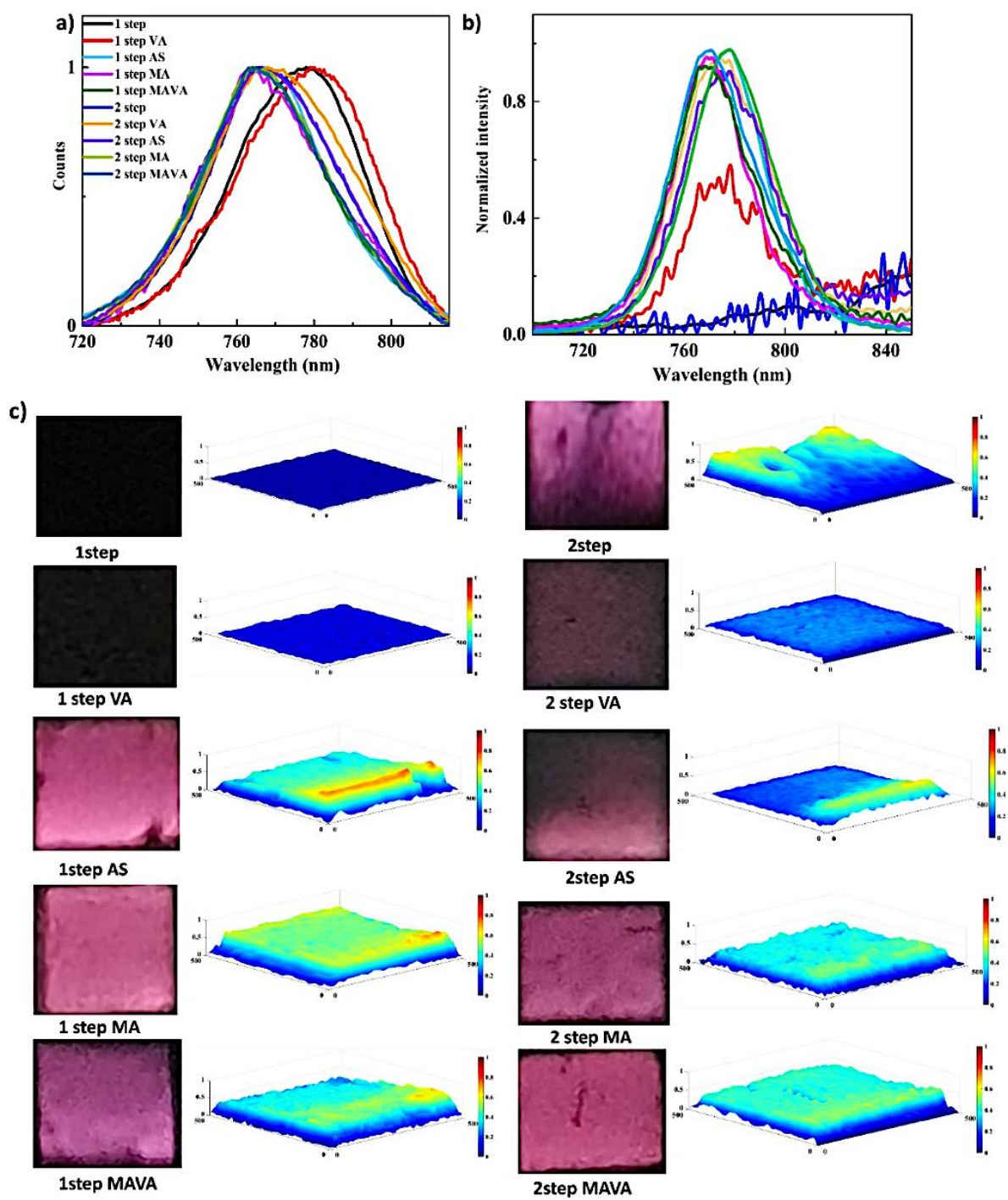

Fig. 4 Recombination studies using PL and EL-based measurements for the perovskite-based devices.

Table 3 Electroluminescence-based parameters and correlation with crystallographic orientations

| Device | Wavelength (nm) | FWHM (nm) | EL Intensity | Injected Voltage (Vi),V | Injected Current (Ii), mA | EQE_EL (%) | Correlation |
|---|---|---|---|---|---|---|---|
| **1 step** | No emission | - | - | - | - | - | Poor film quality |
| **2step** | 783.78 | 71.26 | 0.28 | 2.20 | 20 | 0.09 | Very high current and voltage- remnant $PbI_2$ |
| **1step VA** | 784.90 | 75.37 | 0.011 | 2.5 | 90 | $4\times10^{-4}$ | Very high current and voltage- |
| **2 step VA** | 777.01 | 46.35 | 1.91 | 2.20 | 20 | 0.34 | Very high current and voltage- remnant $PbI_2$ |
| **1step AS** | 777.16 | 36.46 | 0.57 | 1.30 | 6.3 | 0.53 | Strain free- Highest EQE_EL |
| **2 step AS** | 777.20 | 41.81 | 0.39 | 2.2 | 20 | 0.07 | High current-voltage, Perovskite oriented in the $PbI_2$ plane (001) |
| **1stepMA** | 770.56 | 35.99 | 0.27 | 1.3 | 6.5 | 0.25 | Ordered in 110 and 220- high EQE_EL |
| **1step MAVA** | 778.44 | 40.25 | 0.21 | 1.6 | 10 | 0.51 | Highly Ordered in 110 and 220- high EQE_EL |
| **2step MA** | 771.46 | 39.72 | 0.20 | 1.3 | 6 | 0.20 | Strained- lower EQE_EL |
| **2MAVA** | 772.25 | 41.86 | 0.44 | 1.3 | 6.5 | 0.40 | Highly Ordered in 110 and 220- high EQE_EL |

All the high-efficiency devices fabricated involving 1-step and 2-step processes have shown a similar reduction pattern in current density under accelerated ageing conditions for 72 hours (Supporting information Fig. S11).

**Recovery of the water-degraded perovskite films using MAI/IPA dipping and MA vapour treatment**

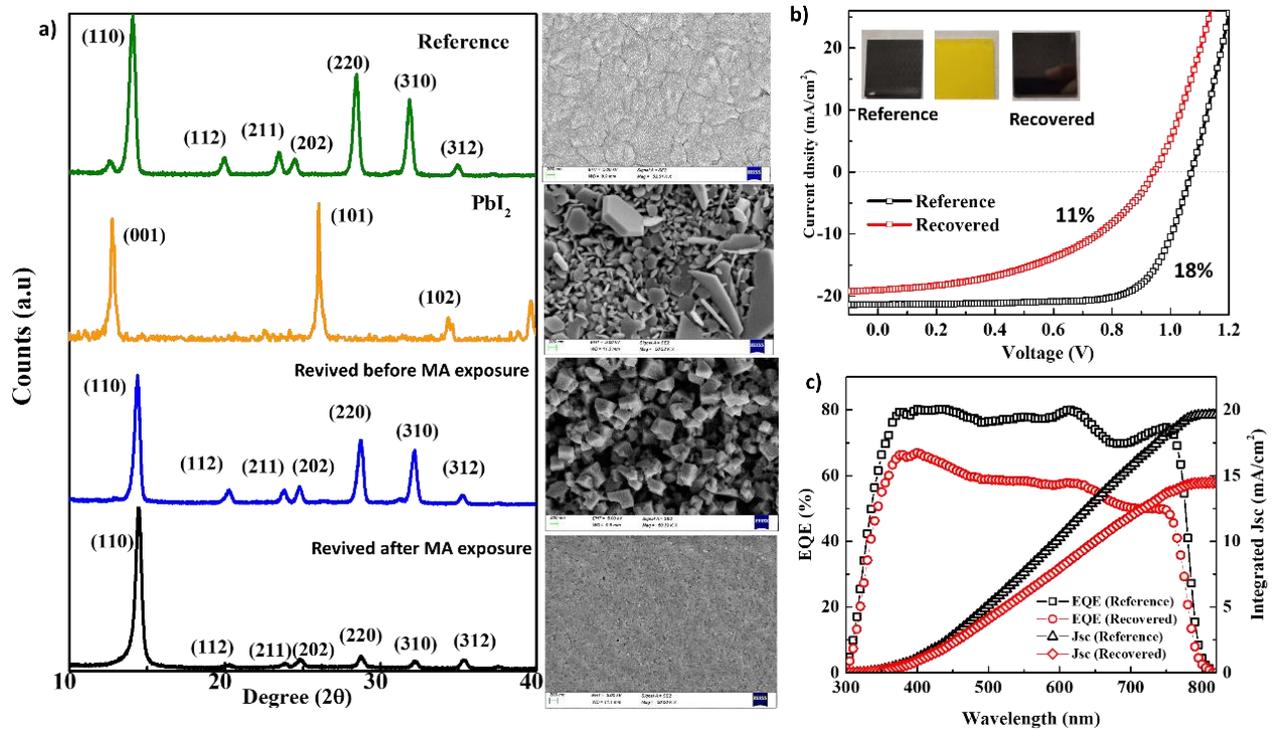

Fig.5 a) GIXRD and morphology of for reference, degraded and revived perovskite films b) Device performance and c) EQE and integrated Jsc for reference and recovered PSCs.

In the context of PSCs and the degradation phenomenon, it is imperative to emphasize the restoration of the deteriorated perovskite material [43]. Literature has indicated that the introduction of amine vapours can partially facilitate the recovery of the perovskite film. Fig. 5a and 5b depict the degradation and subsequent recovery of the perovskite films, as well as the device performance utilizing both reference and rejuvenated perovskite materials. The EQE and integrated Jsc (Fig. 5c) demonstrate a diminished charge carrier effect in the rejuvenated device. It is well-established in the scientific community that the presence of moisture accelerates the degradation of $MAPbI_3$. Water molecules are known to engage in hydrogen bonding interactions with both the lead halide framework and the methylammonium ions. As a result of a negative ΔG, $MAPbI_3$ spontaneously dissociates into MA (aq), $I^-$ (aq) and $PbI_2$ (s) [44]. The $PbI_2$ precipitates on the substrate [45]. Subsequent high-temperature annealing serves to eliminate diffused water molecules, enabling the regeneration of the film through immersion in a methylammonium iodide/isopropanol (MAI/IPA) solution, followed by treatment with methylammonium vapour. It should be noted that lead is soluble in hot water, and there exists a potential risk of lead leaching [45],[46] into the aqueous environment, which can adversely impact the quality of the rejuvenated film. Additionally, the immersion of films in water introduces a high likelihood of contamination which contributed to the lower PCE of the device using revived perovskite film (~11%). Details of the capacitance characteristics and current stability with time of the reference and recovered perovskite are mentioned in the supporting information (Fig. S12a) and S12b).

**Conclusion**

In the conclusion of the manuscript, an extensive investigation was carried out to assess the quality of perovskite films, coated using various processes, with a specific focus on their crystallographic characteristics and thermodynamic aspects. Information regarding the differences in atomic arrangements was provided by GIWAXS. Film quality and disorder, correlated with entropy, were obtained through Raman analysis. Due to the significance of these post-processing techniques in the context of perovskite solar cell recycling, an in-depth analysis encompassing crystallography, defects, morphology, film roughness, and optical phonon modes was undertaken. The effect of these variations in film characteristics was analysed, resulting in device performances of 19.5% for the 1-step AS method and 18.6% for the 2-step MAVA approach. Strong correlations were observed between electroluminescent properties and GIWAXS observations, ultimately revealing that films characterized by enhanced ordering along specific crystallographic planes, such as (110) and (220), were associated with superior device efficiency, higher EQE_EL, and emission occurring at reduced current and voltage levels. The tDOS distribution ranges from 0.22 to 0.35 eV, with a sharp peak for efficient devices and a diffuse pattern for those with poor film quality and high inhomogeneity. A straightforward protocol using MAI/IPA solution and methylamine vapour rejuvenated perovskite films, recovering over 40% device efficiency (Jsc: 19 mA/cm², Voc: 0.94 V), comparable to the reference (Jsc: 22 mA/cm², Voc: 1.07 V). In summary, this comprehensive investigation, focusing on fundamental film properties, luminescent behaviour, and the recycling of perovskite films, is expected to yield valuable insights into the field of perovskite research, contributing significantly to its advancement.

**Conflicts of interest**

No conflicts of interest were reported by the authors.

**Acknowledgement**

This work has been supported by DRDO:DFTM/02/3125/M/13/MNSST-04, 02, March, 2023. The GIWAXS analysis was conducted at the SIRIUS Beamline, SOLEIL Synchrotron facility. The authors would like to express their gratitude to Dr. Phillip Fontaine and Arnaud Hammerle for their assistance with the GIWAXS Experiment. Some parts of the experiments were performed at the Micro-nano Characterization facility (MNCF), CeNSe, IISc Bangalore.